\documentstyle[12pt,epsfig]{article}

\begin{document}
\begin{titlepage}

{\hbox to\hsize{\hfill March 2011 }}

\bigskip \vspace{3\baselineskip}
\begin{center}
{\bf \Large
Top quark forward-backward asymmetry from $SU(N_c)$ color}

\bigskip

\bigskip

\bigskip

{\bf Robert Foot \\}

\bigskip

{ \small \it
ARC Centre of Excellence for Particle Physics at the Terascale,
School of Physics, The University of Melbourne, Victoria 3010, Australia
\\
E-mail: rfoot@unimelb.edu.au
\\}

\bigskip

\bigskip

\bigskip

\bigskip

{\large \bf Abstract}

\end{center}
\noindent
We argue that the $t \bar t$ production asymmetry observed
at the tevatron might be simply explained if the standard $SU(3)_c$ QCD
theory
is extended to $SU(N_c)$ which is spontaneously broken
at a scale just above the weak scale. The extended gauge interactions
amplify
the radiative QCD contribution to the asymmetry and can potentially explain
the observations
if $N_c \stackrel{>}{\sim} 5$. 
This explanation suggests a relatively low $SU(N_c)$ symmetry
breaking scale $\stackrel{<}{\sim} 0.5-1$ TeV.
We check that such a low $SU(N_c)$ symmetry breaking scale 
is consistent with current collider data.
Importantly this scenario predicts an
abundance of striking phenomena
which will be probed at the LHC.
The $SU(N_c)$ model 
also illustrates the idea that a beyond standard model
contribution to the $t \bar t$ asymmetry
might arise primarily via
radiative corrections rather than at tree-level.

\end{titlepage}

\section{Introduction}

The CDF collaboration has observed\cite{cdf} an unexpectedly large
forward-backward $t \bar t$ production asymmetry at the tevatron.
With an integrated luminosity of $5.3\ fb^{-1}$,
the asymmetry is found to vary strongly with the invariant mass
of the $t\bar t$ pair ($M_{t\bar t}$). At the parton level, the
required $t \bar t$ asymmetry in the $t \bar t$ rest frame is observed
to be:
\begin{eqnarray}
A_{t \bar t} &=& 0.475 \pm 0.114 \ {\rm for} \ M_{t\bar t} > 450 \ GeV
\nonumber \\
A_{t \bar t} &=& -0.116 \pm 0.153 \ {\rm for} \ M_{t \bar t} < 450 \ GeV\ .
\end{eqnarray}
These asymmetries can be compared with the next to leading order QCD
prediction of\cite{cdf}
[see also \cite{other}]:
\begin{eqnarray}
A_{t \bar t} &=& 0.088 \pm 0.013 \ {\rm for} \ M_{t\bar t} > 450 \ GeV
\nonumber \\
A_{t \bar t} &=& 0.040 \pm 0.006 \ {\rm for} \ M_{t \bar t} < 450 \ GeV\ .
\end{eqnarray}
The statistical significance of the excess is $3.4 \sigma $ and
represents exciting evidence
for new physics. 
Possible explanations for the asymmetry include s-channel
interference of an exotic color octet gauge boson or t-channel
scalar/vector
exchange, see e.g. ref.\cite{fer,b}
for some recent studies.
All of these explanations
involve an exotic tree-level contribution to the
$t\bar t$ production amplitude and generally have significant
experimental constraints coming from resonance production, 
like-sign top quark production etc. In this paper we wish to explore
the novel possibility that the asymmetry arises predominately
as a radiative effect, analogous to the QCD contribution.

The leading order QCD $q \bar q \to t \bar t$ production asymmetry
arises from two different
reactions\cite{kuhn}. The first involves 
interference of initial state with final
state gluon 
bremsstrahlung. 
The second involves interference betweeen the tree-level $q\bar q \to
t\bar t$ process and its one-loop correction. 
The Feynman diagrams are shown in Figure 1.

\vskip 0.5cm
\centerline{\epsfig{file=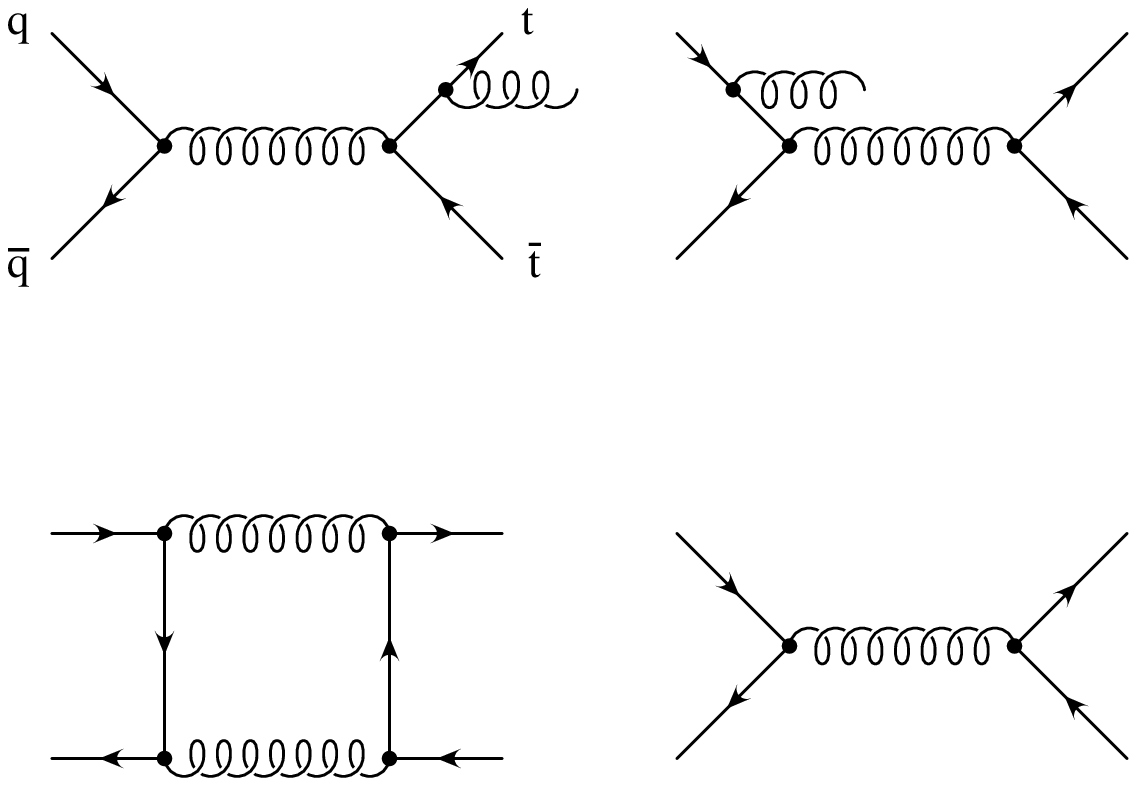, width = 9cm}}
\vskip -0.1cm

\noindent
Figure 1: Interfering $q \bar q \to t \bar t j $ (above) and $q \bar q \to t
\bar t $ (below)
amplitudes.

\vskip 0.7cm

Considering first the asymmetry due to the interference between the
tree-level $q \bar q \to t \bar t$
process and its one-loop correction,
it was found in ref.\cite{kuhn} that the QCD asymmetry
is obtained from the corresponding QED result via the replacement:
\begin{eqnarray}
\alpha_{QED} Q_q Q_t \to {d^2_{abc} \over 16 N_c T_F C_F} \alpha_S = {5
\over 12} \alpha_s
\end{eqnarray}
where $C_F = (N_c^2-1)/2N_c = 4/3$ and $T_F = 1/2$. It is instructive
to make the full $N_c$ dependence explicit. Using $d_{abc}^2 = (N_c^2 -
4)(N_c^2 - 1)/N_c$, we find
that the QCD asymmetry is obtained from the corresponding QED result via
the replacement:
\begin{eqnarray}
\alpha_{QED} Q_q Q_t \to {N_c^2 - 4  \over 4N_c} \alpha_s \ .
\end{eqnarray}
The asymmetry due to the interference of initial state with final
state gluon bremsstrahlung is also proportional to $d_{abc}^2$ and has
the same $N_c$ dependence.
That is, to leading order we have that
\begin{eqnarray}
A_{t\bar t} \propto {N_c^2 - 4 \over 4N_c} \ .
\label{asy6}
\end{eqnarray}
Thus we arrive at the interesting result that the asymmetry increases
rapidly
with increasing $N_c$. This observation suggests that gauge models where
$SU(3)_c$ is
incorporated into a larger $SU(N_c)$ gauge symmetry which is broken
near the weak
scale might provide a suitable candidate for the new physics needed to
explain the $t \bar t$ asymmetry.
Furthermore this illustrates the novel idea that a new physics
contribution to the $t\bar t$ asymmetry could arise primarily via
radiative corrections rather than at tree-level.

\section{The $SU(5)_c$ model - and generalization to $SU(N_c)$}

An example of such an extended gauge model for the strong interactions
is given by
the anomaly free $SU(5)_c \otimes SU(2)_L \otimes U(1)_{Y'}$ gauge model
proposed some
time ago by Hernandez and I\cite{oscar}, and further studied in
ref.\cite{glashow,tom1,tom2,oscar2,hall}.
In the $SU(5)_c$ model, the quarks and leptons of each generation
transform under the gauge symmetry
$SU(5)_c \otimes SU(2)_L \otimes U(1)_{Y'}$ in the anomaly free
representation:
\begin{eqnarray}
f_L &=& \left( \begin{array}{c} \nu \\
e \end{array}\right)_L \sim (1, 2, -1), \ e_R \sim (1, 1, -2), \\
Q_L &=& \left( \begin{array}{c} u \\
d \end{array}\right)_L \sim (5, 2, 1/5), \ u_R \sim (5, 1, 6/5), \ d_R
\sim (5, 1, -4/5)\ .
\end{eqnarray}
Gauge symmetry breaking is accomplished via the vacuum expectation
values (VEVs)
of a scalar $\chi \sim (\overline{10},1,2/5)$ along with the 
usual Higgs scalar,
$\phi \sim (1, 2, +1)$:
\begin{eqnarray}
\langle \chi \rangle = \left( \begin{array}{ccccc} 0 & 0 & 0 & 0 & 0 \\
0 & 0 & 0 & 0 & 0 \\
0 & 0 & 0 & 0 & 0 \\
0 & 0 & 0 & 0 & w \\
0 & 0 & 0 & -w & 0
\end{array} \right) \ ,
\ \ \langle \phi \rangle = \left( \begin{array}{c}
0 \\
u \end{array}\right)\ .
\end{eqnarray}
The $SU(5)_c \otimes U(1)_{Y'}$ gauge symmetry is assumed 
to be spontaneously broken by
the
VEV of $\chi$ at a scale just above the weak scale, which along with
the usual Higgs doublet $\phi$ leads to the symmetry breaking pattern:
\begin{eqnarray}
& SU(5)_c \otimes  SU(2)_L \otimes U(1)_{Y'} &  \nonumber \\
&\downarrow \langle \chi \rangle & \nonumber \\
&SU(3)_c \otimes SU(2)' \otimes SU(2)_L \otimes U(1)_Y &
\nonumber \\
&\downarrow \langle \phi \rangle &
\nonumber \\
&SU(3)_c \otimes SU(2)' \otimes U(1)_Q &
\end{eqnarray}
where $Y = Y' + {T_5 \over \sqrt{15}}$ 
is the linear
combination of $Y'$ and $T_5$ which annihilates $\langle \chi \rangle$
(i.e. $Y\langle \chi \rangle = 0$).
[$T_5$ is the diagonal generator of
$SU(5)_c$ orthogonal
to those of the $SU(3)\otimes SU(2)'$ subgroup, i.e.
$T_5 = \sqrt{{12 \over 5}} diag(1/3,1/3,1/3,-1/2,-1/2)$].

The extra colour degree's of freedom have been called `quirks' and
studied in some
detail by Carlson {\it et al}\cite{hall}. 
These fermions have electric charge $\pm 1/2$
and are confined into integer charged bound states by the unbroken and
asymptotically free $SU(2)'$
gauge interaction [$\Lambda_{SU(2)'} \sim 0.5 \ GeV$]. 
These exotic fermions are given
mass by coupling to $\chi$ via the Yukawa Lagrangian:
\begin{eqnarray}
{\cal L} = \lambda_L \bar Q_L \chi (Q_L)^c + \lambda_R \bar u_R \chi
(d_R)^c \ .
\label{chi2}
\end{eqnarray}
Note that the generation index has been suppressed.
Besides these electroweak invariant masses, there are $SU(5)_c$
invariant mass terms
coming from the usual coupling terms with the standard Higgs doublet,
$\phi$.

In addition to the standard model gauge bosons the model contains a
$Z'_\mu$ and
charged (with electric charge $\pm 1/6$) $W'_\mu$
gauge bosons (from $SU(5)\otimes U(1)_{Y'}/SU(3)\times U(1)_Y$ coset).
The latter transform as
a $(3,2,1/6)$ and $(\bar 3, 2, -1/6)$ 
under the unbroken $SU(3)_c \otimes SU(2)' \otimes U(1)_Q$
gauge group.
These gauge bosons gain masses of\cite{oscar}
\begin{eqnarray} 
M_{Z'}^2 & \simeq & {12 \over 5} g_s^2 w^2\ , \nonumber \\
{M_{W'}^2 \over M_{Z'}^2 } & \simeq & {5 \over 12}\ ,
\end{eqnarray}
where $w$ is the VEV of $\chi$.
The mass eigenstate neutral $\gamma, Z, Z'$ gauge bosons couple to 
fermions as follows:
\begin{eqnarray}
{\cal L} = -e A_\mu J^\mu_{em} - {2e \over \sin 2 \theta_w} Z_\mu
J^\mu_N - g_s Z'_\mu J^\mu_{N'}
\end{eqnarray}
where the currents are given by
\begin{eqnarray}
J^\mu_{em} = \bar \psi Q \gamma^\mu \psi, \
J^\mu_N = \bar \psi N \gamma^\mu \psi, \
J^\mu_{N'} = \bar \psi N' \gamma^\mu \psi \ .
\end{eqnarray}
The summation of fermion fields is implied and the generators $Q, N$ and
$N'$ are given by
\begin{eqnarray}
Q &=& I_3 + {Y' \over 2} + {T_5 \over 2\sqrt{15}}\ , \nonumber \\
N &=& (I_3 - \sin^2 \theta_w Q) + \Delta \ , \nonumber \\
N' &=& {T_5 \over 2} + \Delta'
\label{mon1}
\end{eqnarray}
where $\Delta$ and $\Delta'$ are small corrections given by
\begin{eqnarray}
\Delta \simeq {-e^2 u^2 \sqrt{5/3} T_5 \over 48 g_s^2 w^2 \cos^2 \theta_w}, \
\Delta' \simeq {e^2 Y' \over 2\sqrt{15} g_s^2 \cos^2 \theta_w} \ .
\label{mon2}
\end{eqnarray}
Note that we have defined $\theta_w$ by $\cos^2 \theta_w \equiv
m_W^2/m_Z^2$.

Importantly the scale of $SU(5)_c \otimes U(1)_{Y'}$ symmetry breaking
can be
very low\cite{oscar,glashow,tom1,tom2,oscar2,hall}.
The $Z'$ is very weakly constrained because it couples mainly to hadrons
and its mixing
with the $Z$ is naturally very small. For example, in ref.\cite{glashow}
only the very modest
limit of $m_{Z'} \stackrel{>}{\sim} 100$ GeV was obtained from neutral
current phenomenology.
A more stringent limit of $M_{Z'} > 280$ GeV was obtained from UA2 dijet
data in ref.\cite{tom2}.
In section 4 we will confront this model with more recent dijet data
from the Tevatron where
we show that a low $SU(5)_c$ symmetry breaking scale is still experimentally
allowed, although not without some tension.
The quirk mass is also very weakly constrained. 
A mass limit on the quirks can be obtained from 
measurements of the $Z$ boson width, which limits the mass of the
quirks to be greater than $43$ GeV. In addition, flavour changing neutral
current processes,
such as the radiative contribution to $K^0 - \bar K^0$ mass mixing
constrain the quirk masses to be
roughly degenerate or the corresponding CKM-type matrix to be almost
diagonal\cite{hall}.

The case of larger (odd) $N_c$ is a straightforward generalization to
the $SU(5)_c$ case, but
requires a more complex Higgs sector to achieve the required symmetry
breaking\footnote{
The case of even $N_c$ is possible, although the simplest such theories
run into phenomenological
problems\cite{fh2}.}. To break the symmetry in a
phenomenologically consistent way in $SU(N_c)$ with $N_c > 5$ will
require several scalar multiplets.
For example, an $SU(7)_c \otimes SU(2)_L
\otimes U(1)_{y'}$ model
can be constructed by placing the fermions into the anomaly-free
representation:
\begin{eqnarray}
f_L &=& \left( \begin{array}{c} \nu \\
e \end{array}\right)_L \sim (1, 2, -1), \ e_R \sim (1, 1, -2), \\
Q_L &=& \left( \begin{array}{c} u \\
d \end{array}\right)_L \sim (7, 2, 1/7), \ u_R \sim (7, 1, 8/7), \ d_R
\sim (5, 1, -6/7) \ .
\end{eqnarray}
The $SU(7)_c \otimes U(1)_{Y'}$ can be spontaneously broken by
the
VEVs of two scalars, $\chi_1, \chi_2 \sim (\overline{21},1,2/7)$ at 
a scale just above the weak scale.
These VEVs are given by:
\begin{eqnarray}
\langle \chi_1 \rangle = \left( \begin{array}{ccccccc} 0 & 0 & 0 & 0 & 0 &
0 & 0 \\
0 & 0 & 0 & 0 & 0 & 0 & 0\\
0 & 0 & 0 & 0 & 0 & 0 & 0\\
0 & 0 & 0 & 0 & 0 & 0 & 0\\
0 & 0 & 0 & 0 & 0 & 0 & 0\\
0 & 0 & 0 & 0 & 0 & 0 & w_1\\
0 & 0 & 0 & 0 & 0 & -w_1 & 0
\end{array} \right) \ , \ 
\langle \chi_2 \rangle = \left( \begin{array}{ccccccc} 0 & 0 & 0 & 0 & 0 &
0 & 0 \\
0 & 0 & 0 & 0 & 0 & 0 & 0\\
0 & 0 & 0 & 0 & 0 & 0 & 0\\
0 & 0 & 0 & 0 & w_2 & 0 & 0\\
0 & 0 & 0 & -w_2 & 0 & 0 & 0\\
0 & 0 & 0 & 0 & 0 & 0 & 0\\
0 & 0 & 0 & 0 & 0 & 0 & 0
\end{array} \right) \ .
\end{eqnarray}
These VEVs will give the exotic quirk degree's of freedom electroweak
invariant masses, analogously
to the $SU(5)_c$ case [Eq.(\ref{chi2})].
These VEVs along with $\langle \phi \rangle$
lead to the symmetry breaking pattern:
\begin{eqnarray}
& SU(7)_c \otimes  SU(2)_L \otimes U(1)_{Y'} &  \nonumber \\
&\downarrow \langle \chi_1 \rangle, \langle \chi_2 \rangle & \nonumber
\\
&SU(3)_c \otimes SU(2)' \otimes SU(2)' \otimes SU(2)_L \otimes U(1)_Y &
\nonumber \\
&\downarrow \langle \phi \rangle &
\nonumber \\
&SU(3)_c \otimes SU(2)' \otimes SU(2)' \otimes U(1)_Q &
\end{eqnarray}
where $Y = Y' + {2 \over \sqrt{42}}T_7$. Here  $T_7$ denotes 
the $SU(7)$ diagonal generator:
\begin{eqnarray}
T_7 = \sqrt{{24 \over 7}} diag(1/3,1/3,1/3,-1/4,-1/4,-1/4,-1/4) \ .
\end{eqnarray}
The $SU(7)_c$ model will contain two $Z'$ gauge bosons, $Z'_1, Z'_2$.
These gauge bosons couple to the generators
\begin{eqnarray}
Z'_1 & : & \ \cos\theta' {T_7 \over 2} + \sin\theta' {T_x\over 2}
\nonumber \\
Z'_2 & : & \  - \sin\theta' {T_7 \over 2} + \cos\theta' {T_x \over 2}
\end{eqnarray}
where $T_7$ is defined above and $T_x =
\sqrt{2}diag(0,0,0,-1/2,-1/2,1/2,1/2)$, and $\theta'$
is obtained by diagonalizing the $Z'$ gauge boson mass matrix:
\begin{eqnarray}
\left( \begin{array}{cc}
{6\over 7}g_s^2 (w_1^2 + w_2^2)& \sqrt{{12\over 7}} g_s^2 (w_1^2 -
w^2_2)\nonumber \\
\sqrt{{12\over 7}} g_s^2 (w_1^2 - w^2_2) & 2g_s^2 (w_1^2 + w_2^2)
\end{array}\right)\ .
\end{eqnarray}
That is, $\tan 2\theta' = {\sqrt{21} \over 2} \left[ {w_2^2 - w_1^2
\over w_2^2 + w_1^2} \right]$.
For $SU(7)_c$ and the case of larger $N_c$, other symmetry
breaking patterns are possible depending on the scalar content
of the theory.

\section{Top quark forward-backward asymmetry in $SU(N_c)$ color
theories}

The QCD contributions to the $t \bar t$ forward-backward asymmetry
feature
infrared (IR) divergences. The contribution due
to gluon bremsstrahlung can be separated into hard and soft 
gluon emission, and the IR divergence from the latter cancelling
the IR divergence from the box diagam contribution to the
asymmetry\cite{other,kuhn}.
Calculations show that the hard gluon bremsstrahlung
contribution to $A_{t\bar t}$ is negative, 
while the soft gluon plus virtual part
is positive\cite{other,kuhn}.  It turns
out that the soft gluon bremsstrahlung + virtual box diagram part is about
twice the magnitude of the hard bremstrahlung part, so that the  
predicted QCD asymmetry is positive.

In the extended $SU(N_c)$ color gauge models, the $t\bar t$
asymmetry should be approximately
the same as that predicted in the
standard model at energies below the scale of $SU(N_c)$ symmetry
breaking, which
we take here for illustration as a single scale, $m$.
The radiative processes involving exotic $W',Z'$ gauge bosons and quirks
are suppressed by powers of $\sqrt{\hat{s}}/m$.
Thus, things are clear in the $m \gg \sqrt{\hat{s}}$ limit - the
standard model calculation of the $t\bar t$ asymmetry results.
Let us now examine the $m \ll \sqrt{\hat{s}}$ limit. In this limit
we expect the $SU(N_c)$ symmetry to be a good approximation and
provided that $T$ quirks 
(we denote the quirk partner of the corresponding quark with
an uppercase letter)
decay into t-quarks 
\footnote{The part of the asymmetry due to the
interference of initial and final state bremsstrahlung will contain
$q\bar q \to t W' T$ final states (as well as $q\bar q \to t \bar t g$
final states).
The $T$ quirks can decay into top quarks: $T \to t U \bar u$
via a virtual $W'$, assuming that the $U$ quirk is the lightest quirk.},
then the
$t \bar t$ asymmetery should follow from Eq.(\ref{asy6}).
If the region $M_{t \bar t} > 450$ GeV is sufficiently high
that the $SU(N_c)$ limit becomes useful, then we can predict
from Eq.(\ref{asy6}) a $t \bar t$ forward-backward asymmetry of:
\begin{eqnarray}
A_{t \bar t} & \approx & 0.22 \ {\rm for } \ N_c = 5 \nonumber \\
A_{t \bar t} & \approx  & 0.34 \ {\rm for } \ N_c = 7 \nonumber \\
A_{t \bar t} & \approx & 0.45 \ {\rm for } \ N_c = 9 \nonumber \\
A_{t \bar t} & \approx & 0.56 \ {\rm for } \ N_c = 11 \ \ etc.
\label{56}
\end{eqnarray}
In the realistic case we might have $\sqrt{\hat{s}} \sim m$
and that $T$ quirks might not decay predominately into t-quarks,
in any case the bremsstrahlung contribution can potentially
be kinematically supressed. 
In this case, the exotic contribution to the $t \bar t$ asymmetry
should arise predominately from the interference of the
tree-level $q \bar q \to t \bar t$ amplitude with the 
virtual box diagram amplitude. 
It seems that
the size of this contribution can be even larger than
that suggested by
the $SU(N_c)$ symmetry estimate, Eq.(\ref{56}), 
since that estimate
includes the hard bremstrahlung contribution, which is negative in
sign.
It is also possible that the exotic box diagram contributions
are smaller than the $SU(N_c)$ estimate, Eq.(\ref{56})
or even that it has the wrong sign (although it is unlikely that it has
the wrong sign for all ranges of parameters).
For completeness one also needs to calculate the box diagram
contribution with gauge bosons replaced by the $\chi$ scalar/Goldstone
bosons (in 't Hooft-Feynman gauge). 
The latter could dominate if the quirks are heavier than
the gauge bosons.
In this paper we will be content with pointing out the novel
possibility that the top quark forward-backward asymmetry
might be radiatively induced with $SU(N_c)$ models as a promising
example.
Obviously if the CDF anomaly is confirmed by future data then more
detailed calculations will be necessary to pin down the
$m_{quirk}/m_{Z',W'}/N_c$ parameter space allowed by the experiments.

Observe that the total top quark production cross section 
in $SU(N_c)$ color models
is not expected to be
significantly ($\stackrel{<}{\sim} 15\%$) modified from the
standard model prediction. This is because the main effect of the new
physics is to 
increase the radiative contribution which generates the
asymmetry via interference with the QCD contribution.
The proportion of $T$ quirks produced is relatively small.

If $SU(N_c)$ color gauge theory is the origin of the $t\bar t$ 
asymmetry then
the scale of new physics 
is likely to be low (unless $N_c$ is very big), and we thus expect
this explanation to
be tested in the future at the LHC experiment. However in the meantime
we need to check
that the low $SU(N_c)$ symmetry breaking scale is compatible with
existing data.

\section{Tevatron limits on $SU(5)_c$ [$SU(N_c)$] symmetry breaking
scale}

Data from the Tevatron might constrain the $SU(5)_c$ symmetry breaking
scale.
The most promising signature is expected to be the effects
of the $Z'$ since this particle can be produced on resonance if
its mass is low\cite{tom1,tom2}.
The $Z'$ will manifest itself primarily as a bump in the dijet invariant
mass distribution.
No such bump is seen, leading to strong constraints on possible $Z'$
coupling to quarks.
For example, the CDF collaboration\cite{cdfdijets} are able to exclude
an axigluon with mass
$260 < m < 1250$ GeV.
However, it turns out that the $SU(5)_c$ $Z'$ couples to each quark
color with coupling
$g_s/\sqrt{15}$. The $1/\sqrt{15}$ factor significantly suppresses the
cross section c.f. axigluons.
We have computed the cross section for $ p \bar p \to Z' \to 2 jets$. 
This is obtained by calculating
the cross section for the parton subprocess $q \bar q \to Z' \to  q' \bar
q'$ and then folding this in
with the parton distribution functions in the usual way.
We use the 2008MSTW NLO parton distribution functions in
our numerical work\cite{mstw}.
The cross section is multiplied by the standard $K$ factors
\begin{eqnarray}
K_i = 1 + {8\pi \alpha_s (q^2) \over 9}, \ K_f = 1 + {\alpha_s (q^2)
\over \pi}
\end{eqnarray}
to incorporate the QCD corrections in the initial and final states.
We have also applied a rapidity cut of $|y| < 1$ for each of
the two jets.
Our results are shown in
figure 2. The calculation assumes that the quirks are light, so that
they contribute to the total width of the $Z'$
but do not contribute to the dijet signal\footnote{Quirk pairs can
annihilate into jets, but at a lower
invariant mass, where the dijet constraints are much weaker.}.
Also shown in the figure is the corresponding CDF $95\%$ C.L. upper
limit\cite{cdfdijets}.

\vskip 1cm
\centerline{\epsfig{file=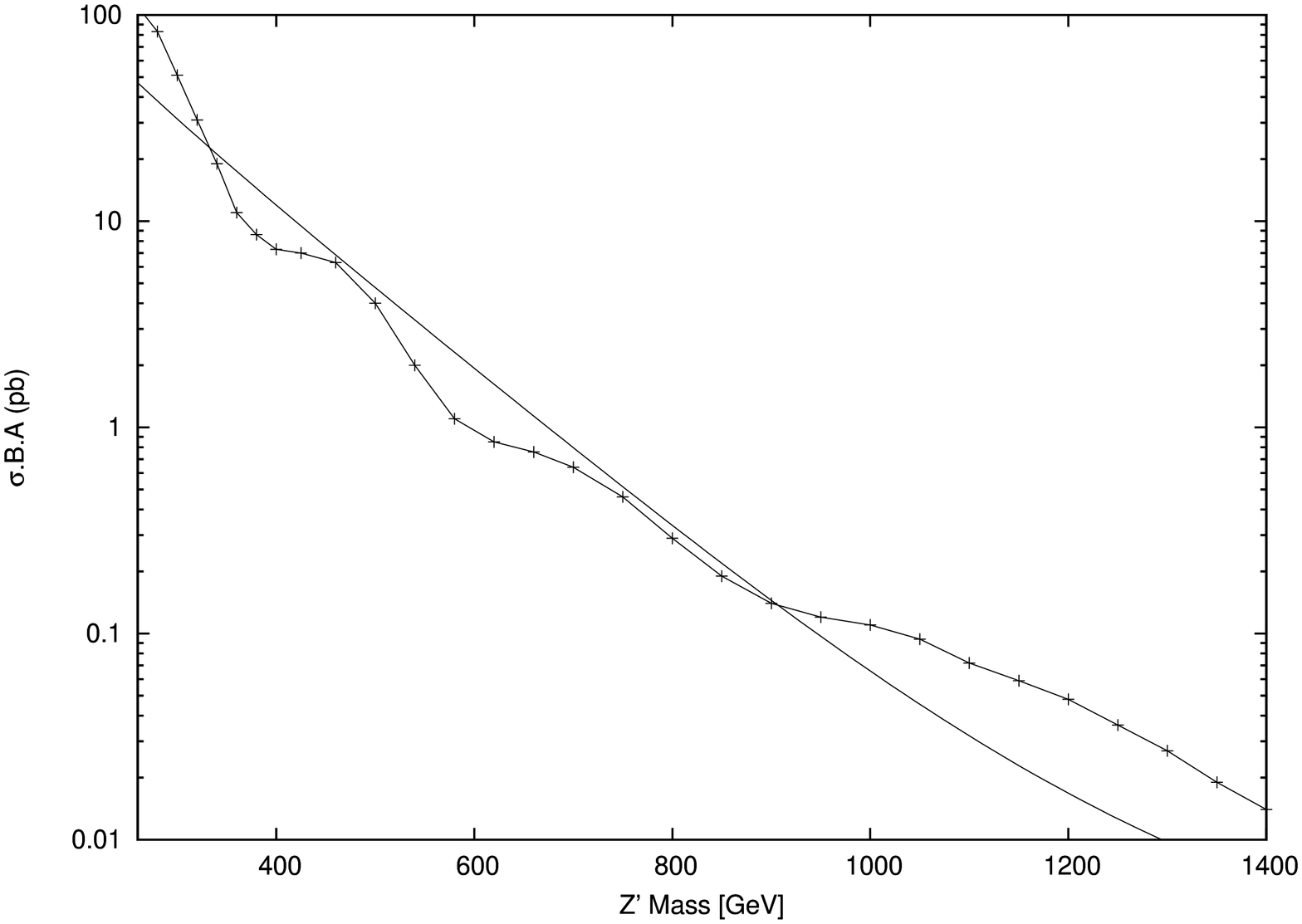, angle = 270, width = 12.6cm}}
\vskip 0.4cm
\noindent
Figure 2: The $Z'$ particle production cross section times branching
fraction into dijets times the acceptence for
both jets to have $|y| < 1$. Also shown, is
the corresponding CDF $95\%$ C.L. upper limit.

\vskip 1cm


As the figure shows, there is some tension between a low $SU(5)_c$
breaking scale, which disfavours the
model but is not quite strong enough to conclusively exclude it. In the
case of $SU(7)_c$ and
larger color groups, having multiple $Z's$ might help evade the CDF
dijet constraint by effectively smoothing
out the dijet bump.
The $Z'$ can also decay leptonically via a small
$Y'$ coupling [from the $\Delta'$ part in Eqs.(\ref{mon1}, \ref{mon2})].
However the branching fraction to leptons is very small,
for example we find that $Br(Z' \to \mu \bar \mu) \simeq 1.4\times
10^{-4}$. In fact this branching fraction is sufficiently small
to evade tevatron limits\cite{cdfmuon} on resonances decaying into muon
pairs.

The tevatron might potentially constrain the quirk masses.
Quirk pairs can be produced via $\gamma, Z, Z'$ exchange.
While heavy quirks can decay into lighter ones, the lightest pair of
quirks are expected to hadronize (assuming
the lightest quirk is lighter than the $W'$ or $SU(2)'$ colored scalar
components).
As discussed in some detail in ref.\cite{hall},
this quirk pair can be viewed as a highly excited quirkonium state. 
Deexcitation occurs via numerous emission of $SU(2)'$ glueballs, dubed
'hueballs' in ref.\cite{hall}.
Note that the hueball emission will lead to substantial
missing energy but only small missing transverse momentum.
The quirkonium will end up in the ground state where it 
will decay via virtual $\gamma, Z$ into standard model particles.

The quirk production cross section times branching ratio 
into muons 
at the tevatron has been calculated
in ref.\cite{hall}. They obtain a cross section times branching ratio
into muon pairs of
about $10 fb$ for a quirk mass of 100 GeV for left-handed quirks, and
much weaker limits for right-handed
quirks. This is well below current limits obtained by the CDF
collaboration\cite{cdfmuon}.
Thus, we conclude that a low $SU(N_c)$ symmetry breaking scale of a
few hundred GeV is phenomenologically viable,
and consequently the $SU(N_c)$ explanation of the top quark
forward-backward asymmetry is possible.

Observe that the $SU(N_c)$ explanation of the $t \bar t$
forward-backward asymmetry
would imply a similarly sized asymmetry for $b \bar b$ quark production. 
It has
been argued in ref.\cite{b,c} that
such an asymmetry, although challenging, could likely be measured at the
tevatron.
If this is the case then this will further test the $SU(N_c)$ model.

The CDF collaboration has also recently found an anomalously
high number of boosted jets ($p_T > 400$ GeV for the leading
jet) with invariant mass near the top quark mass\cite{cdf2}.
The anomaly could be due to an underestimation of background
but might also be a signal for new physics.
Such a signal can be explicable in $SU(N_c)$ models.
For instance,
if $m_{Z'} \sim 0.8-1$ TeV, then $Z' \to t\bar t$ decays will
generally lead to a boosted top. Furthermore, from figure 2 we see that
the cross section for hadronically decaying
$Z'$ is of order $0.1$ pb, which implies a cross section
to hadronically decaying top pairs of about 10 fb.
This value is roughly consistent with the value necessary
to explain the CDF data if it is due to new physics\cite{cdf2,perez}.
The CDF collaboration did not observe any anomaly in the
boosted jet plus missing energy channel (corresponding to
one of the top quarks decaying
semileptonically). However
it has been argued in ref.\cite{perez2}  that the semileptonic channel
is less sensitive than the fully hadronic decay channel.

The LHC should be able to probe the $SU(N_c)$ 
explanation for the $t \bar t$
asymmetry.
The quirks, $W',Z'$ gauge bosons and also the exotic colored 
$\chi$ scalar can be produced at the LHC, leading
to dijet
signals. We expect, however, that at least several $fb^{-1}$ of data might
need to be accumulated before this
explanation of the $t\bar t$ asymmetry can be put to the test.
We leave detailed predictions for the LHC for future work.

\section{Conclusion}

In conclusion, we have pointed out that the $t \bar t$ forward-backward
asymmetry observed
at the tevatron might be simply explained if the standard $SU(3)_c$ QCD
theory
is extended to $SU(N_c)$ which is spontaneously broken
at a scale just above the weak scale. The extended gauge interactions
amplify
the QCD radiative contribution to the asymmetry and 
can potentially explain the
observations
if $N_c \stackrel{>}{\sim} 5$. 
This explanation suggests a relatively low $SU(N_c)$ symmetry
breaking scale $\stackrel{<}{\sim} 0.5-1$ TeV, and we have checked
that this is compatible with recent collider data.
Importantly this scenario predicts an
abundance of new physics
just above the weak scale,
which will be probed in the future at the LHC.

\subsection*{Acknowledgements}

The author thanks Archil Kobakhidze for useful conversations and
to G. Rodrigo for some helpful correspondence.
This work was supported in part by the Australian Research Council.

\end{document}